\documentclass[
reprint,
superscriptaddress,
amsmath,amssymb,
aps,
prl,
]{revtex4-2}

\usepackage{graphicx}% Include figure files
\usepackage{dcolumn}% Align table columns on decimal point
\usepackage{bm}% bold math
\usepackage[colorlinks=true,allcolors=blue]{hyperref}% add hypertext capabilities
\begin{document}

\preprint{APS/123-QED}

\title{Spiral waves speed up cell cycle oscillations in the frog cytoplasm}% Force line breaks with \\

\author{Daniel Cebri{\'a}n-Lacasa}
\affiliation{Laboratory of Dynamics in Biological Systems, Department of Cellular and Molecular Medicine, KU Leuven, Herestraat, 49, Leuven, Belgium}
\affiliation{Centre for Engineering Biology, University of Edinburgh, Edinburgh EH9 3BF, United Kingdom}
\author{Liliana Pi{\~n}eros}
\affiliation{Laboratory of Dynamics in Biological Systems, Department of Cellular and Molecular Medicine, KU Leuven, Herestraat, 49, Leuven, Belgium}
\author{Arno Vanderbeke}
\affiliation{Laboratory of Dynamics in Biological Systems, Department of Cellular and Molecular Medicine, KU Leuven, Herestraat, 49, Leuven, Belgium}
\author{Daniel Ruiz-Reyn{\'e}s}
\affiliation{Laboratory of Dynamics in Biological Systems, Department of Cellular and Molecular Medicine, KU Leuven, Herestraat, 49, Leuven, Belgium}
\affiliation{IFISC (CSIC-UIB). Instituto de Física Interdisciplinar y Sistemas Complejos, E-07122 Palma de Mallorca, Spain}
\author{Thibeau Wouters}
\affiliation{Laboratory of Dynamics in Biological Systems, Department of Cellular and Molecular Medicine, KU Leuven, Herestraat, 49, Leuven, Belgium}
\affiliation{Institute for Gravitational and Subatomic Physics (GRASP), Utrecht University, Princetonplein 1, 3584 CC Utrecht, The Netherlands}
\affiliation{Nikhef, Science Park 105, 1098 XG Amsterdam, The Netherlands}
\author{Andrew B. Goryachev}
\affiliation{Centre for Engineering Biology, University of Edinburgh, Edinburgh EH9 3BF, United Kingdom}
\author{Nikita Frolov}
\affiliation{Laboratory of Dynamics in Biological Systems, Department of Cellular and Molecular Medicine, KU Leuven, Herestraat, 49, Leuven, Belgium}
\affiliation{Authors contributed equally} 
\author{Lendert Gelens}
\affiliation{Laboratory of Dynamics in Biological Systems, Department of Cellular and Molecular Medicine, KU Leuven, Herestraat, 49, Leuven, Belgium}
\affiliation{Authors contributed equally}

\date{\today}

\begin{abstract}
Spiral waves are a well-known phenomenon in excitable media, playing critical roles in biological systems such as cardiac tissues, where they are involved in arrhythmias, and in slime molds, where they guide collective cell migration. However, their presence in the cytoplasm of cells has not been reported to date. 
In this study, we present the observation of spiral waves in a \textit{Xenopus laevis} frog egg extract reconstituting periodic cell cycle transitions. We find that the emergence of these spiral waves accelerates the cell division cycle nearly twofold. Using two distinct computational models, we demonstrate that this behavior arises from generic principles and is driven primarily by time-scale separation in the cell cycle oscillator. Additionally, we investigate the interplay between these spiral waves and the more commonly observed target pattern waves in the frog cytoplasm, providing new insights into their dynamic interactions.
%Spiral waves have been proven to play essential roles in biology, particularly in cardiac systems and cortical dynamics. Nevertheless, to date, no one has reported the presence of spirals in cytoplasmic dynamics. In this work, we present the first-ever observation of spiral waves in \textit{Xenopus laevis} cell-free extract and analyze their influence on the period of microtubule signaling and nuclear signaling. This analysis provides insights into the effect these dynamics might have on the pace of cell division. Our data indicate a reduction in the period when spiral waves are present. Subsequently, we used two different computational models to demonstrate that this behavior is generic and primarily depends on time-scale separation, disappearing when the separation does. Furthermore, we use this result to estimate the time-scale separation of the cytoplasmic regulatory network.
\end{abstract}


\maketitle
Dynamic processes in living systems are highly organized in both space and time, enabling the precise coordination of essential biological functions. Understanding the mechanisms underlying spatiotemporal pattern formation is a fundamental challenge at the interface of physics and biology. Self-sustained oscillations, such as those driving neuronal dynamics, cell division cycles, heartbeats, and circadian rhythms, are ubiquitous and critical for life \cite{Novak2008,Beta2017}. Disruptions in these periodic oscillations are linked to severe diseases, including cancer and heart arrhythmias \cite{Pertsov1993,Davidenko1995,Panfilov1996,Panfilov2004,PANFILOV20191}. Recently, the spatial coordination of these oscillations over long distances via inter- and intracellular waves has emerged as a key focus of research, as wave-driven coordination is critical for maintaining proper cellular function and developmental processes \cite{Beta2017,Gelens2014, Deneke2018}.

Living systems are characterized by a constant input of energy and dissipation, thus being nonequilibrium systems.  In physics, wave patterns in nonequilibrium systems have long been studied and they emerge from the interplay between local dynamical processes and diffusive transport \cite{Fife1976,Keener1980,TYSON1988327,MERON19921,Gelens2014,Beta2017,Deneke2018}. In oscillatory and excitable media, waves often propagate at constant speeds, forming continuous families of periodic wave trains. In two dimensions, common patterns include expanding circular target waves, driven by fast-oscillating pacemaker regions \cite{Rombouts2020}, and rotating spiral waves, which arise from defects or inhomogeneities.
Spiral waves are a ubiquitous feature of both chemical and biological systems \cite{KEENER1986307,cebrian2024six}, where they can serve important physiological functions. For instance, in starved Dictyostelium cells, spiral waves of cAMP coordinate collective movement and slime mold formation \cite{TYSON1989193,Lechleiter1991,Siegert1995,Ford2023}. In cardiology, spiral waves are linked to heart arrhythmias, making them a key focus of research \cite{Pertsov1993,Davidenko1995,Panfilov1996,Panfilov2004,PANFILOV20191}.
% Occasionally, spiral calcium waves are observed in frog oocytes \cite{Lechleiter1991}.
\begin{figure}[t!]
    \centering
    \includegraphics[width=0.49\textwidth]{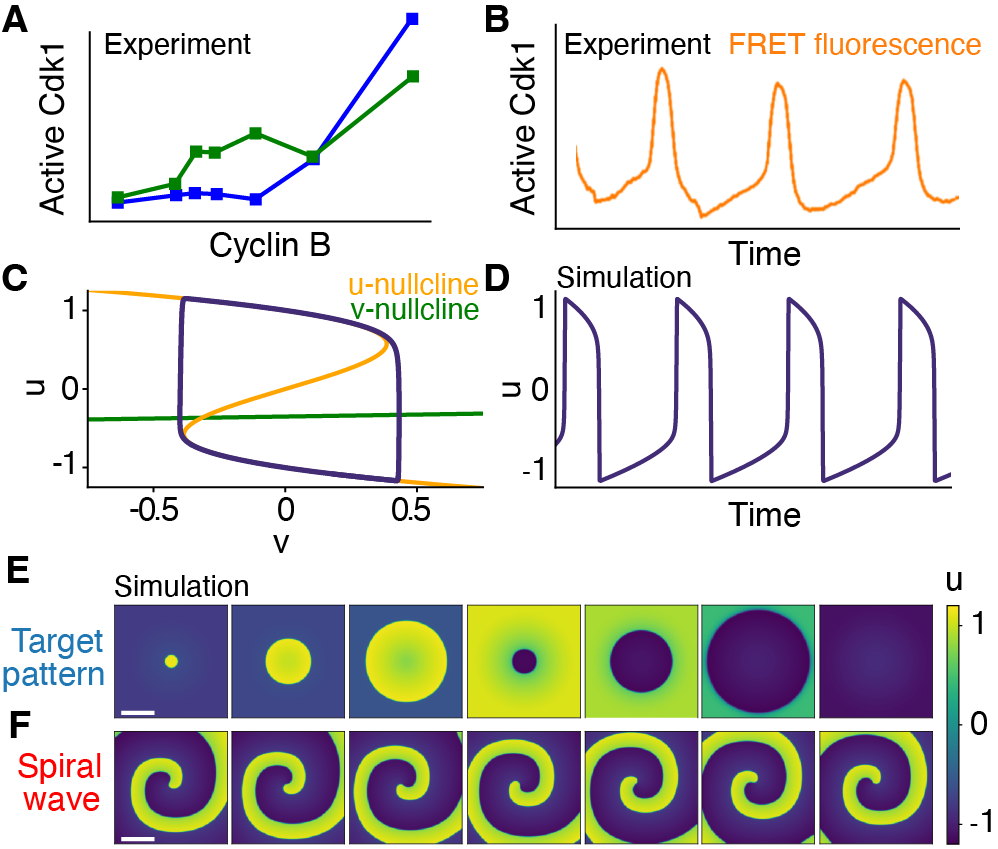}
    \caption{\textbf{Target patterns and spiral waves in excitable media (simulations).} \textbf{A} Bistability of active Cdk1 as a function of cyclin B concentration (data from \cite{Pomerening2003-ef}). \textbf{B} Cell cycle oscillations in frog egg extract measured using a Cdk1 FRET sensor (\cite{Maryu2022-ul}). \textbf{C-D} Relaxation oscillations in the FHN model are shown in phase space and time series. \textbf{E} Target patterns in the FHN model from a central pacemaker ($b_\text{M}=0.5$). \textbf{F} Spiral waves in the FHN model from a phase defect in initial conditions. Scale bars: 75 px.}
    \label{fig:Figure1}
\end{figure}
\begin{figure*}[t!]
    \centering
    \includegraphics[width=\textwidth]{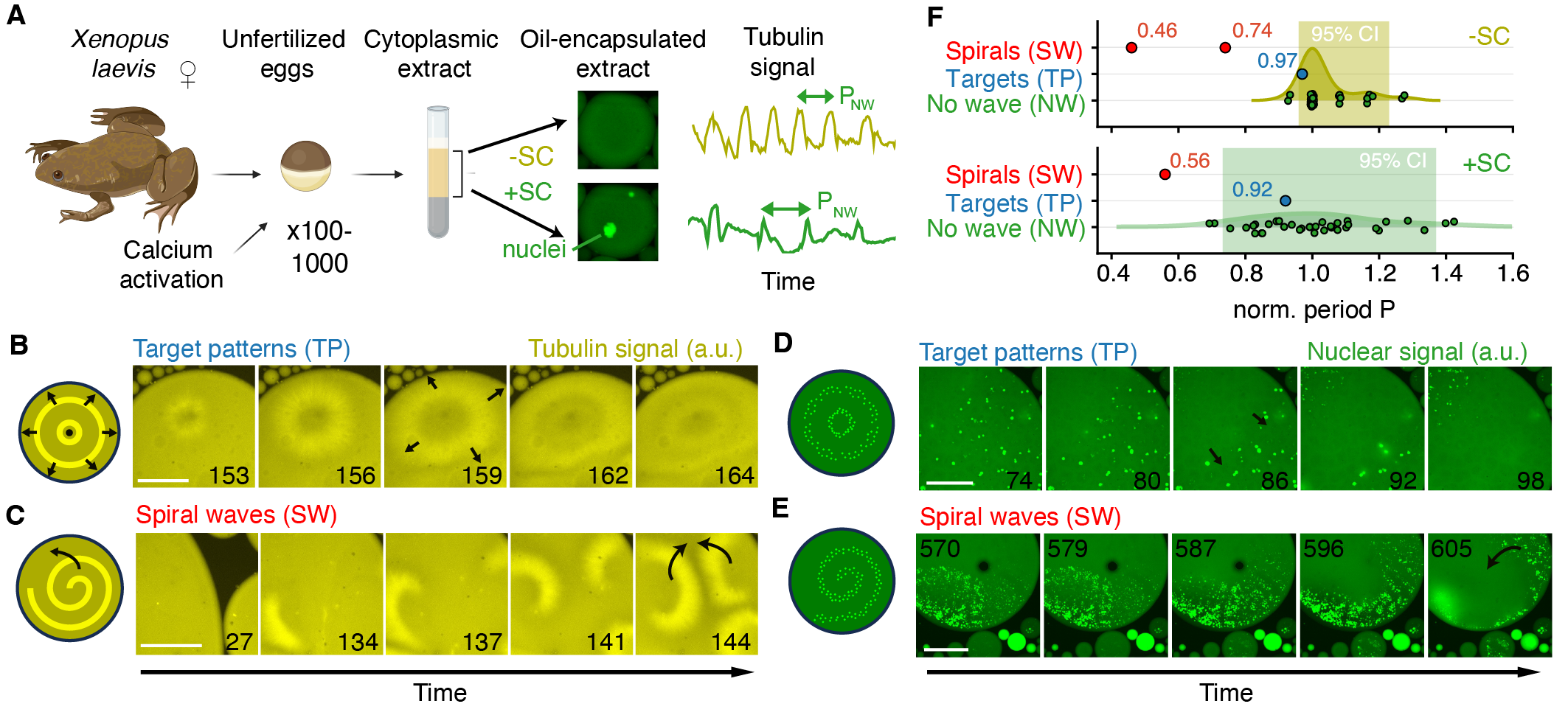}
    \caption{\textbf{Target patterns and spiral waves in frog egg extracts (experiments).}
\textbf{A} Schematic of the experimental setup to reconstruct cell cycle oscillations in oil-encapsulated droplets of \textit{Xenopus} egg extracts. \textbf{B-C} Target patterns (TP) and spiral waves (SW) were observed in droplets using a tubulin reporter. \textbf{D-E} TP and SW observed with a nuclear reporter (NLS-GFP).  \textbf{F} Normalized oscillation periods ($P_{NW}$) compared across droplets without waves ($N$=45: without SC, $N$=36: with SC) and droplets with TP or SW. Panel \textbf{B}: Scale bar $200\mu$m. Panel \textbf{C}: Scale bar $350\mu$m. Panels \textbf{D, E}: Scale bar $300\mu$m. Time stamps are in minutes.}
    \label{fig:Figure2}
\end{figure*}
Another remarkable example of physiological control through traveling waves is the regulation of the early cleavages in large embryos such as the \textit{Xenopus laevis} frog (about 1mm in diameter) \cite{Gelens2014,Chang2013,Puls2024}. It starts with a calcium wave triggered by sperm entry, preventing refertilization by sweeping across the egg in a target pattern \cite{fontanilla1998characterization}. After fertilization, surface contraction waves—ripples in the cell cortex—precede cytokinesis \cite{hara1980cytoplasmic}. In vitro studies link these ripples to biochemical target pattern waves of cell cycle regulators' activity originating at the nucleus and spreading through the cytoplasm to coordinate division \cite{Chang2013,Nolet2020,Afanzar2020,Puls2024}.
One major regulator, Cyclin-dependent kinase 1 (Cdk1), drives the early cell cycle by periodically alternating its activity, regularly switching between two branches of a bistable response to cyclins \cite{Pomerening2003-ef} (Fig.~\ref{fig:Figure1}A-B). This relaxation oscillator \cite{Novak1993,parra2023cell,Maryu2022-ul} resembles the FitzHugh-Nagumo (FHN) model, featuring an S-shaped nullcline and time-scale separation ($\varepsilon$) \cite{FitzHugh1961,Nagumo1962,cebrian2024six} (Fig.~\ref{fig:Figure1}C-D):

\begin{equation}
    \begin{array}{ @{} l  @{} }
        u_t=D\Delta u-u^3+u-v, \\[\jot]
        v_t=D\Delta v+\varepsilon(u-bv+a).
    \end{array}
\end{equation}

The FHN model, commonly used to study traveling waves, is known to support both target patterns and spiral waves \cite{cebrian2024six} (Fig.~\ref{fig:Figure1}E-F). 
Similarly, both wave types also appear in dynamical cell cycle models (see Supplemental Material), highlighting they are generic features of excitable systems. However, spiral waves have not been experimentally observed in the cytoplasm of early frog embryos until now.
In this work, we fill this gap by reporting spiral waves in frog egg extracts that surprisingly decrease the cell cycle period by up to two-fold. Through numerical modeling, we demonstrate such a period reduction to be a general property of spiral waves in the presence of sufficient time-scale separation.

%Using frog egg extracts, we report spiral waves that decrease the cell cycle period by up to two-fold, consistent with computational predictions.

%Fig.\ \ref{fig:Figure1} illustrates the formation of both waves in the FitzHugh-Nagumo (FHN) model, a generic 2D model for excitable media with time scale separation ($\epsilon$) and cubic nonlinearity \cite{FitzHugh1961,Nagumo1962,cebrian2024six}:

\begin{figure*}[t!]
    \centering
    \includegraphics[width=1.\textwidth]{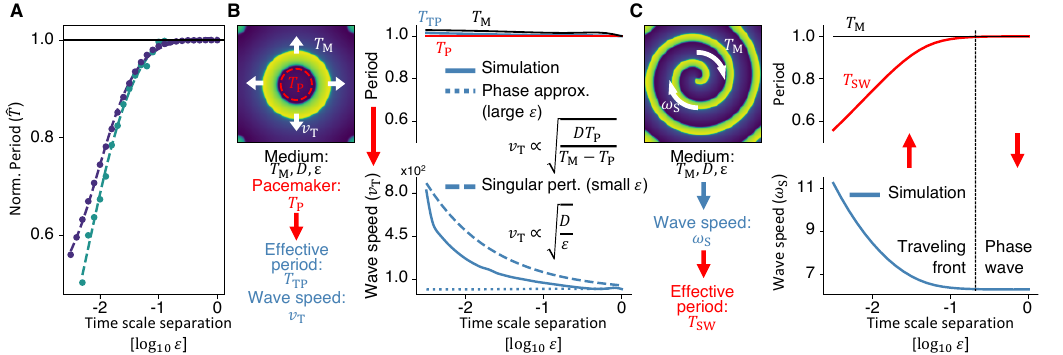}
    \caption{
    \textbf{Spiral waves reduce the oscillation period depending on the time-scale separation.}
\textbf{A.} A sweep across $\varepsilon$ shows a strong relationship between period reduction and time-scale separation in both the FHN model and a cell cycle model \cite{parra2023cell}, well-described by a Hill function fit.
\textbf{B.} Target patterns generated by a pacemaker region, with simulations showing the dependence of the oscillation period and wave speed $v_T$ on the time-scale separation $\varepsilon$. Two analytical approximations for wave speed are included.
\textbf{C.} Spiral waves induced by a central phase defect, with simulations illustrating the dependence of the oscillation period and angular velocity $\omega_S$ on the time-scale separation $\varepsilon$.}
    \label{fig:Figure3}
\end{figure*}

Our in vitro experiments were conducted as follows. We prepared frog egg extract from hundreds of \textit{Xenopus laevis} eggs (Fig.~\ref{fig:Figure2}A) by activating them with calcium ionophore to initiate cell cycle oscillations, then crushing through centrifugation and extracting the cytoplasmic fraction. This extract was encapsulated in surfactant-stabilized oil droplets, producing hundreds of droplets per experiment with diameters normally ranging from $\sim$70 to 300$\mu$m, but reaching up to $\sim$1mm, mimicking cell sizes during early \textit{X. laevis} cleavage divisions. Dynamics were visualized using fluorescence time-lapse microscopy (see Supplemental Material). Fig.~\ref{fig:Figure2}A shows $\sim$25-minute periodic oscillations of fluorescently-labeled tubulin intensity in one of the extract droplets, reflecting self-(dis)assembly of the mitotic spindle/aster apparatus driven by periodic changes of Cdk1 activity. This process mimics the cell cycle and matches its duration in cleaving embryos~\cite{Pineros2024.07.28.605512,anderson2017desynchronizing}.

Typically, in sufficiently large droplets ($\sim$1mm diameter), we observed target pattern waves (Fig.~\ref{fig:Figure2}B; Movies 1--3). However, in 5 experiments, we found 7 droplets exhibiting spiral wave dynamics, including coexisting counter-rotating spirals (Fig.~\ref{fig:Figure2}C; Movies 4--10). In one of the experiments, spiral waves were found in bulk cell-free extract (Movie 11). We hypothesize that underlying Cdk1 activity waves drive the observed waves of tubulin polymerization. On the other hand, Cdk1 can potentially stimulate microtubule branching~\cite{johmura2011regulation}, exhibiting intrinsic wave-like activity \cite{Ishihara_eLife,Rinaldin2024-wz} that can interfere with the Cdk1 activity waves. To exclude the interference effect, we repeated our experiment, supplementing the extract with nuclear material -- demembranated sperm chromatin (SC) -- to enable self-organization of nuclei, visualized with GFP-NLS. In this setup, the alternation of interphase (nuclear presence) and mitosis (nuclear envelope breakdown and GFP-NLS diffusion) is synchronized with tubulin fluorescence (Fig.~\ref{fig:Figure2}A; Supplemental Material), confirming nuclei as an alternative reliable readout of cell cycle oscillations. Consistent with earlier experiments \cite{Chang2013,Nolet2020,Afanzar2020,Puls2024}, we observed target patterns of Cdk1 activity propagating through the cytoplasm at $\approx0.41\mu$m/s (Fig.~\ref{fig:Figure2}D, Movie 12)  and, strikingly, the emergence of spiral waves in larger droplets (Fig.~\ref{fig:Figure2}E, Movie 13). Together, these experiments support the link of the observed wave patterns to Cdk1 activity waves.

%The GFP-NLS signal was synchronized with tubulin fluorescence, confirming the latter as a reliable readout of cell cycle oscillations (Fig.\ref{fig:Figure2}A; Supp. Fig. ***).

%In larger droplets, we again observed target patterns (Fig.\ref{fig:Figure2}D). Strikingly, spiral waves also emerged in the presence of nuclei, further supporting their link to Cdk1 activity waves (Fig.~\ref{fig:Figure2}E).

\begin{figure*}[t]
    \centering
    \includegraphics[width=\textwidth]{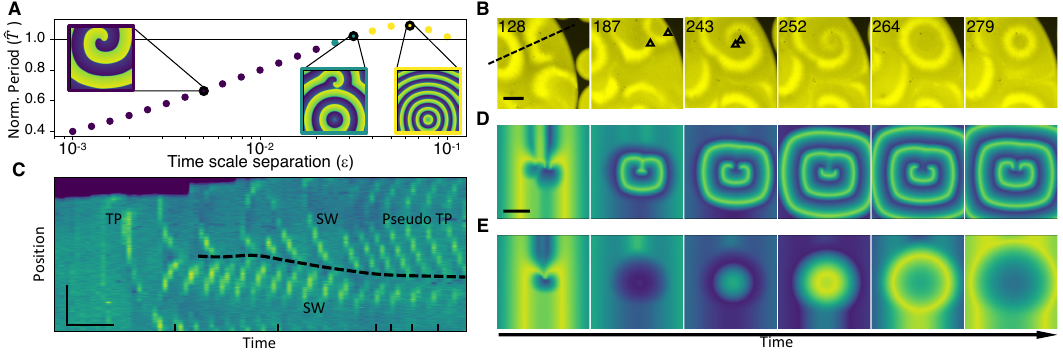}
    \caption{\textbf{Competition between spirals and target patterns.} 
    \textbf{A.} The plot shows how, in the presence of spirals and target patterns, the one with the fastest periodicity dominates the system. Corresponding kymographs for each point are provided in Supplemental Material. Exceptional parameters: $(a,b_\text{P},b_\text{M})=(0,0,1)$.
    \textbf{B, C.} Frames of a droplet exhibiting spirals in which not only the target patterns do not disappear, but the period of the target patterns seems to adapt to one of the spiral waves (time in min). Kymograph performed on the black dashed line of the first frame. Scale bars: $150 \mu$m and $30$ min
    \textbf{D, E.} Frames of simulated spirals in which two defects are close in space and lead to seemingly target patterns. In particular, panel E illustrates how the defects merge and circular waves propagate from the defect. A double tip defect explains the violation of the results shown in A and agrees with the double tip observed in B. Scale bar: $100$ px.}
    \label{fig:Figure4}
\end{figure*}

In both experimental conditions, spiral waves were observed in large droplets mainly under two conditions: i) when two smaller droplets merged (Fig.~\ref{fig:Figure2}C; Movies 4,8), introducing phase defects; and ii) when an air bubble was present (Fig.~\ref{fig:Figure2}E, Movie 5,13), bending a traveling wave upon collision with the obstacle. These observations align with established mechanisms of spiral wave generation in excitable media \cite{KEENER1986307,Zykov2018,Glass_2022,Babloyantz_spirals,Panfilov_spirals}.
Visual inspection suggested shorter periods for spirals compared to target patterns. However, noisy conditions and temporal resolution limited reliable period measurements to three experiments: two with the tubulin marker and one with the nuclei marker. This resulted in 5 large droplets exhibiting spatial dynamics (3 had spiral waves, and 2 had target patterns). Using wavelet transforms, we identified oscillation regions in frequency and time, averaging the frequency spectra to determine mean periods (see Supplemental Material). Comparing periods from larger droplets with spirals (SW), target patterns (TP), and smaller droplets with no waves (NW) (Fig.~\ref{fig:Figure2}F), we normalized by the median $\tilde{T}=T_{\text{SW}}/T_{\text{NW}}$ for each experiment and aggregated the data. While target patterns corresponded closely to the median NW period, SW periods deviated significantly, confirming the observed faster oscillations and reduced periodicity in the presence of spiral waves (Movies 14--16).

To determine whether the period reduction is an intrinsic property of our system or a general feature of excitable systems, we turned to computational modeling. We used the FHN model, widely used to study excitable systems \cite{cebrian2024six} and cell cycle oscillations \cite{Gelens2014,Nolet2020}.
First, simulations across all combinations of FHN model parameters (see Supplemental Material) revealed that the period reduction relative to the uncoupled system without waves ($\hat{T}=T_\text{SW}/T_\text{NW}$) was primarily influenced by the time-scale separation $\varepsilon$, with minimal dependence on $a$ and $b$. Strong time-scale separation also enabled self-excitation of spirals, expanding the oscillatory parameter space. Additionally, large system size was essential for spiral induction, particularly at small $\varepsilon$ values.
Next, we analyzed the oscillation period of both the FHN model and a known cell cycle model (see Supplemental Material) as $\varepsilon$ varied (Fig~\ref{fig:Figure3}A). Both showed similar period reductions for decreasing $\varepsilon$, with $\hat{T}$ saturating at $1$ as $\varepsilon$ increased. Fitting the Hill function to these response curves, we estimated the time-scale separation in the cytoplasmic extract: $\bar{\varepsilon}_\text{CC}\approx0.008$ and $\bar{\varepsilon}_\text{FHN}\approx0.004$.
%was driving the region to an enlargement of the oscillatory region, yielding to preserved local oscillations on combinations of parameters that do not oscillate naturally if the system  is oscillatory. This characteristic vanishes as the time-scale separation is reduced, since excitability as a feature of systems with large time-scale separation. Finally, we also observed that the system must be large enough so spirals can be induced, particularly, in the case of large time-scale separation. 

We hypothesized that the observed period reduction with respect to time-scale separation could be due to the differences in how target and spiral waves are generated and propagate through a medium.
%In previous work, we quantified the wave speed of pacemaker-generated waves in 1D, corresponding to target patterns in 2D \cite{Rombouts2020, Rombouts_PhysRevE.104.014220}. 
 Fig.~\ref{fig:Figure3}B shows that for target patterns, the effective period $T_{\text{TP}}$ remains nearly constant, lying between the pacemaker period $T_\text{P}$ and the medium period $T_\text{M}$. However, wave speed increases up to 10-fold as $\varepsilon$ decreases, consistent with our previous works~\cite{Rombouts2020, Rombouts_PhysRevE.104.014220}. Using singular perturbation theory \cite{TYSON1988327, Rombouts_PhysRevE.104.014220}, we find that wave speed scales as $\sqrt{D / \varepsilon}$ for small $\varepsilon$, where target patterns behave like traveling fronts, and as $\sqrt{D T_\text{P} / (T_\text{M}-T_\text{P})}$ for large $\varepsilon$, where pacemaker region is significant and period differences are minor.
 Together, these results indicate a weak effect of speed on the periodicity of target pattern waves.
 %Thus, pacemaker and medium properties set the effective period and wave speed for target patterns.
For spiral waves, the dynamics differ. Without a pacemaker, the system has a single intrinsic period $T_\text{M}$, and spiral waves arise from heterogeneous initial conditions, such as defects. Their wave speed (in this case, angular velocity $\omega_s$) also depends on $\varepsilon$, but the periodicity is determined by the time required for the spiral to complete one turn $2\pi/\omega_s$. Consequently, the period saturates at $T_\text{M}$ for larger $\varepsilon$ when the medium's natural frequency exceeds the one set by the spiral (Fig.~\ref{fig:Figure3}C), leading to a transition from traveling fronts to phase waves.

%To explain why the period reduction occurs we used the FHN to plot the linear velocity and normalized period as a function of the time scale separation in target patterns (see Fig.~\ref{fig:Figure3}B, C and D), and the angular velocity and normalized period in spiral waves (see Fig.~\ref{fig:Figure3}E, F and G). While the former showed no correlation between the normalized period and the linear velocity, the latter exhibited clear correlation among the angular velocity and the normalized period. This is an indicator that in the case of target patterns the waves do not drive the period of the system, as they are generated at a fixed pace related with the natural periodicity of the system, while they do in the case of spiral waves, as the period is set by the amount of time that the wave needs to do a complete loop. Furthermore, this implies that when spiral waves are present the period saturates at one because the natural oscillatory frequency is faster than the one set by the traveling waves, indicating that the observed phenomena is the phase waves rather than traveling waves after the saturation occurs.

To further test the mechanism of spiral wave formation, we performed additional simulations. In Supplemental Material, we show how the reduction in period depends on the diffusion coefficient. Consistent with our hypothesis, the period decreases further as the diffusion increases, due to the corresponding rise in the wave speed. Since diffusion of enzymatic complexes such as Cdk1 may be slower due to binding kinetics and cytoplasmic heterogeneity, the actual separation on a time scale could be stronger than our estimate of $\bar{\varepsilon}_\text{CC}\approx0.008$. We examined the transition from traveling waves to phase waves, observing changes not only in the period but also in wave morphology (see Supplemental Material). For large time-scale separations, the system forms Archimedean-like spirals, while smaller separations give logarithmic-like spirals. Analyzing oscillation phases, we find that reduced times in bottleneck regions result from self-excitation. The amplitude of oscillations in $v$ is significantly smaller for spiral waves compared to target patterns and uniform oscillations. This occurs because the system bypasses the natural oscillation bottleneck, instead relying on incoming waves for excitation, effectively skipping phase delays.

Finally, to understand why spirals remain stable coexisting with target patterns in our experiments, we investigated the competition between these waves. 
We simulated a system where a spiral and a target pattern were induced on opposite sides using a phase defect and a pacemaker, respectively. In all simulations, the wave with the fastest periodicity ultimately dominated (Fig.~\ref{fig:Figure4}A; Supplemental Material) \cite{PhysRevLett.79.2907}. On the contrary, experimental movies revealed that spirals could be replaced by coexisting target patterns that shared the same (faster) periodicity as spiral waves (Fig.~\ref{fig:Figure4}B-C).
We considered two hypotheses for this behavior. First, target patterns might adapt to the periodicity of spiral waves. However, as shown in Fig.~\ref{fig:Figure3}, target periodicity is governed by the system's natural oscillatory frequency, meaning that adaptation would require changes in intrinsic parameters. Alternatively, we hypothesized that closely spaced defects could merge, emitting waves resembling a target pattern. To test this, we initialized two defects, observing that their emitted waves combined into a single propagating wave (Fig.~\ref{fig:Figure4}D). When the defect tips were brought sufficiently close, the two spirals merged, producing target-like waves and eliminating visible phase defects (Fig.~\ref{fig:Figure4}E). Indeed, Fig.~\ref{fig:Figure4}B does suggest that two spirals merge (see black triangles).

In conclusion, we demonstrated the emergence of spiral waves in the cytoplasm of early \textit{Xenopus laevis} embryos, driven by periodic Cdk1 activity. Using frog egg extracts encapsulated in droplets, we observed spiral waves alongside target patterns, with spiral waves significantly reducing the cell cycle period. Simulations based on the FitzHugh-Nagumo model confirmed that this period reduction is a robust property of excitable systems with strong time-scale separation. Our findings highlight the critical interplay between local oscillatory dynamics, diffusive transport, and initial conditions in shaping wave patterns, providing new insights into the self-organization of the cell cycle. This observation of cytoplasmic spiral waves complements previously identified spiral waves in the \textit{Xenopus} oocyte and embryo, including spiral patterns observed in fertilization calcium waves \cite{Lechleiter1991} and on the cell cortex \cite{bement2015activator, bement2024patterning}.

The observation of spiral waves in the cytoplasmic extract opens new avenues for investigating their functional roles in embryonic development. Unlike target patterns, spiral waves can shorten the cell cycle period, potentially altering early cleavage dynamics. Experimentally, spiral waves could be actively induced in the \textit{Xenopus} cytoplasm by introducing controlled phase defects or geometric perturbations, such as obstacles or droplet fusion. These approaches would allow in vitro studies to explore spiral wave properties in detail, including their dependence on system size, diffusion coefficients, and Cdk1 activity levels.
In vivo, spiral waves may have physiological consequences, similar to their role in cardiac dynamics, where they are associated with arrhythmias. Although spatial chaos has theoretically been suggested to disrupt early embryonic development, experimental evidence indicates that this is unlikely to occur under physiological conditions \cite{GELENS2015892}. Investigating the effects of spiral waves on multicellular organization, division timing, and embryonic robustness could provide insights into their developmental significance. Additionally, understanding the interactions between spiral waves and natural target patterns could reveal mechanisms governing transitions between wave types, with broader implications for excitable media and self-organizing biological systems.\\

\textit{Data availability.} The numerical codes to reproduce the figures in this study are openly available in \textsc{GitLab} [Upcoming], and as an archived repository in RDR by KU Leuven [Upcoming]. The experimental data is openly available on RDR [Upcoming].\\

\textit{Acknowledgements.}
L.G. acknowledges funding by the KU Leuven Research Fund (grant number C14/23/130) and the Research-Foundation Flanders (FWO, grant number G074321N). L.P. acknowledges funding by the Research-Foundation Flanders (FWO, personal fellowship - grant number 11I4521N).
D.R.-R. is supported by the Ministry of Universities through the ``Pla de Recuperacio, Transformació i Resiliencia'' and by the EU (NextGenerationEU), together with the Universitat de les Illes Balears.

\nocite{*} %Show the non-cited entries

\bibliography{references}% Produces the bibliography via BibTeX.

\begin{thebibliography}{46}%
\makeatletter
\providecommand \@ifxundefined [1]{%
 \@ifx{#1\undefined}
}%
\providecommand \@ifnum [1]{%
 \ifnum #1\expandafter \@firstoftwo
 \else \expandafter \@secondoftwo
 \fi
}%
\providecommand \@ifx [1]{%
 \ifx #1\expandafter \@firstoftwo
 \else \expandafter \@secondoftwo
 \fi
}%
\providecommand \natexlab [1]{#1}%
\providecommand \enquote  [1]{``#1''}%
\providecommand \bibnamefont  [1]{#1}%
\providecommand \bibfnamefont [1]{#1}%
\providecommand \citenamefont [1]{#1}%
\providecommand \href@noop [0]{\@secondoftwo}%
\providecommand \href [0]{\begingroup \@sanitize@url \@href}%
\providecommand \@href[1]{\@@startlink{#1}\@@href}%
\providecommand \@@href[1]{\endgroup#1\@@endlink}%
\providecommand \@sanitize@url [0]{\catcode `\\12\catcode `\$12\catcode `\&12\catcode `\#12\catcode `\^12\catcode `\_12\catcode `\%12\relax}%
\providecommand \@@startlink[1]{}%
\providecommand \@@endlink[0]{}%
\providecommand \url  [0]{\begingroup\@sanitize@url \@url }%
\providecommand \@url [1]{\endgroup\@href {#1}{\urlprefix }}%
\providecommand \urlprefix  [0]{URL }%
\providecommand \Eprint [0]{\href }%
\providecommand \doibase [0]{https://doi.org/}%
\providecommand \selectlanguage [0]{\@gobble}%
\providecommand \bibinfo  [0]{\@secondoftwo}%
\providecommand \bibfield  [0]{\@secondoftwo}%
\providecommand \translation [1]{[#1]}%
\providecommand \BibitemOpen [0]{}%
\providecommand \bibitemStop [0]{}%
\providecommand \bibitemNoStop [0]{.\EOS\space}%
\providecommand \EOS [0]{\spacefactor3000\relax}%
\providecommand \BibitemShut  [1]{\csname bibitem#1\endcsname}%
\let\auto@bib@innerbib\@empty
%</preamble>
\bibitem [{\citenamefont {Novák}\ and\ \citenamefont {Tyson}(2008)}]{Novak2008}%
  \BibitemOpen
  \bibfield  {author} {\bibinfo {author} {\bibfnamefont {B.}~\bibnamefont {Novák}}\ and\ \bibinfo {author} {\bibfnamefont {J.~J.}\ \bibnamefont {Tyson}},\ }\bibfield  {title} {\bibinfo {title} {Design principles of biochemical oscillators},\ }\href {https://doi.org/10.1038/nrm2530} {\bibfield  {journal} {\bibinfo  {journal} {Nature Reviews Molecular Cell Biology}\ }\textbf {\bibinfo {volume} {9}},\ \bibinfo {pages} {981} (\bibinfo {year} {2008})}\BibitemShut {NoStop}%
\bibitem [{\citenamefont {Beta}\ and\ \citenamefont {Kruse}(2017)}]{Beta2017}%
  \BibitemOpen
  \bibfield  {author} {\bibinfo {author} {\bibfnamefont {C.}~\bibnamefont {Beta}}\ and\ \bibinfo {author} {\bibfnamefont {K.}~\bibnamefont {Kruse}},\ }\bibfield  {title} {\bibinfo {title} {Intracellular oscillations and waves},\ }\href {https://doi.org/10.1146/annurev-conmatphys-031016-025210} {\bibfield  {journal} {\bibinfo  {journal} {Annual Review of Condensed Matter Physics}\ }\textbf {\bibinfo {volume} {8}},\ \bibinfo {pages} {239} (\bibinfo {year} {2017})}\BibitemShut {NoStop}%
\bibitem [{\citenamefont {Pertsov}\ \emph {et~al.}(1993)\citenamefont {Pertsov}, \citenamefont {Davidenko}, \citenamefont {Salomonsz}, \citenamefont {Baxter},\ and\ \citenamefont {Jalife}}]{Pertsov1993}%
  \BibitemOpen
  \bibfield  {author} {\bibinfo {author} {\bibfnamefont {A.~M.}\ \bibnamefont {Pertsov}}, \bibinfo {author} {\bibfnamefont {J.~M.}\ \bibnamefont {Davidenko}}, \bibinfo {author} {\bibfnamefont {R.}~\bibnamefont {Salomonsz}}, \bibinfo {author} {\bibfnamefont {W.~T.}\ \bibnamefont {Baxter}},\ and\ \bibinfo {author} {\bibfnamefont {J.}~\bibnamefont {Jalife}},\ }\bibfield  {title} {\bibinfo {title} {Spiral waves of excitation underlie reentrant activity in isolated cardiac muscle.},\ }\href {https://doi.org/10.1161/01.RES.72.3.631} {\bibfield  {journal} {\bibinfo  {journal} {Circulation Research}\ }\textbf {\bibinfo {volume} {72}},\ \bibinfo {pages} {631} (\bibinfo {year} {1993})}\BibitemShut {NoStop}%
\bibitem [{\citenamefont {Davidenko}\ \emph {et~al.}(1995)\citenamefont {Davidenko}, \citenamefont {Salomonsz}, \citenamefont {Pertsov}, \citenamefont {Baxter},\ and\ \citenamefont {Jalife}}]{Davidenko1995}%
  \BibitemOpen
  \bibfield  {author} {\bibinfo {author} {\bibfnamefont {J.~M.}\ \bibnamefont {Davidenko}}, \bibinfo {author} {\bibfnamefont {R.}~\bibnamefont {Salomonsz}}, \bibinfo {author} {\bibfnamefont {A.~M.}\ \bibnamefont {Pertsov}}, \bibinfo {author} {\bibfnamefont {W.~T.}\ \bibnamefont {Baxter}},\ and\ \bibinfo {author} {\bibfnamefont {J.}~\bibnamefont {Jalife}},\ }\bibfield  {title} {\bibinfo {title} {Effects of pacing on stationary reentrant activity: theoretical and experimental study},\ }\href {https://doi.org/10.1161/01.RES.77.6.1166} {\bibfield  {journal} {\bibinfo  {journal} {Circulation research}\ }\textbf {\bibinfo {volume} {77}},\ \bibinfo {pages} {1166} (\bibinfo {year} {1995})}\BibitemShut {NoStop}%
\bibitem [{\citenamefont {Aliev}\ and\ \citenamefont {Panfilov}(1996)}]{Panfilov1996}%
  \BibitemOpen
  \bibfield  {author} {\bibinfo {author} {\bibfnamefont {R.~R.}\ \bibnamefont {Aliev}}\ and\ \bibinfo {author} {\bibfnamefont {A.~V.}\ \bibnamefont {Panfilov}},\ }\bibfield  {title} {\bibinfo {title} {A simple two-variable model of cardiac excitation},\ }\href {https://doi.org/10.1016/0960-0779(95)00089-5} {\bibfield  {journal} {\bibinfo  {journal} {Chaos, Solitons \& Fractals}\ }\textbf {\bibinfo {volume} {7}},\ \bibinfo {pages} {293} (\bibinfo {year} {1996})}\BibitemShut {NoStop}%
\bibitem [{\citenamefont {Nash}\ and\ \citenamefont {Panfilov}(2004)}]{Panfilov2004}%
  \BibitemOpen
  \bibfield  {author} {\bibinfo {author} {\bibfnamefont {M.~P.}\ \bibnamefont {Nash}}\ and\ \bibinfo {author} {\bibfnamefont {A.~V.}\ \bibnamefont {Panfilov}},\ }\bibfield  {title} {\bibinfo {title} {Electromechanical model of excitable tissue to study reentrant cardiac arrhythmias},\ }\href {https://doi.org/10.1016/j.pbiomolbio.2004.01.016} {\bibfield  {journal} {\bibinfo  {journal} {Progress in Biophysics and Molecular Biology}\ }\textbf {\bibinfo {volume} {85}},\ \bibinfo {pages} {501} (\bibinfo {year} {2004})}\BibitemShut {NoStop}%
\bibitem [{\citenamefont {Panfilov}\ \emph {et~al.}(2019)\citenamefont {Panfilov}, \citenamefont {Dierckx},\ and\ \citenamefont {Volpert}}]{PANFILOV20191}%
  \BibitemOpen
  \bibfield  {author} {\bibinfo {author} {\bibfnamefont {A.}~\bibnamefont {Panfilov}}, \bibinfo {author} {\bibfnamefont {H.}~\bibnamefont {Dierckx}},\ and\ \bibinfo {author} {\bibfnamefont {V.}~\bibnamefont {Volpert}},\ }\bibfield  {title} {\bibinfo {title} {Reaction--diffusion waves in cardiovascular diseases},\ }\href {https://doi.org/10.1016/j.physd.2019.04.001} {\bibfield  {journal} {\bibinfo  {journal} {Physica D: Nonlinear Phenomena}\ }\textbf {\bibinfo {volume} {399}},\ \bibinfo {pages} {1} (\bibinfo {year} {2019})}\BibitemShut {NoStop}%
\bibitem [{\citenamefont {Gelens}\ \emph {et~al.}(2014)\citenamefont {Gelens}, \citenamefont {Anderson},\ and\ \citenamefont {Ferrell}}]{Gelens2014}%
  \BibitemOpen
  \bibfield  {author} {\bibinfo {author} {\bibfnamefont {L.}~\bibnamefont {Gelens}}, \bibinfo {author} {\bibfnamefont {G.~A.}\ \bibnamefont {Anderson}},\ and\ \bibinfo {author} {\bibfnamefont {J.~E.}\ \bibnamefont {Ferrell}},\ }\bibfield  {title} {\bibinfo {title} {Spatial trigger waves: positive feedback gets you a long way},\ }\href {https://doi.org/10.1091/mbc.e14-08-1306} {\bibfield  {journal} {\bibinfo  {journal} {Molecular Biology of the Cell}\ }\textbf {\bibinfo {volume} {25}},\ \bibinfo {pages} {3486} (\bibinfo {year} {2014})},\ \bibinfo {note} {pMID: 25368427}\BibitemShut {NoStop}%
\bibitem [{\citenamefont {Deneke}\ and\ \citenamefont {Di~Talia}(2018)}]{Deneke2018}%
  \BibitemOpen
  \bibfield  {author} {\bibinfo {author} {\bibfnamefont {V.~E.}\ \bibnamefont {Deneke}}\ and\ \bibinfo {author} {\bibfnamefont {S.}~\bibnamefont {Di~Talia}},\ }\bibfield  {title} {\bibinfo {title} {Chemical waves in cell and developmental biology},\ }\href {https://doi.org/10.1083/jcb.201701158} {\bibfield  {journal} {\bibinfo  {journal} {Journal of Cell Biology}\ }\textbf {\bibinfo {volume} {217}},\ \bibinfo {pages} {1193} (\bibinfo {year} {2018})}\BibitemShut {NoStop}%
\bibitem [{\citenamefont {Fife}(1976)}]{Fife1976}%
  \BibitemOpen
  \bibfield  {author} {\bibinfo {author} {\bibfnamefont {P.~C.}\ \bibnamefont {Fife}},\ }\bibfield  {title} {\bibinfo {title} {Pattern formation in reacting and diffusing systems},\ }\href {https://doi.org/10.1063/1.432246} {\bibfield  {journal} {\bibinfo  {journal} {The Journal of Chemical Physics}\ }\textbf {\bibinfo {volume} {64}},\ \bibinfo {pages} {554} (\bibinfo {year} {1976})}\BibitemShut {NoStop}%
\bibitem [{\citenamefont {Keener}(1980)}]{Keener1980}%
  \BibitemOpen
  \bibfield  {author} {\bibinfo {author} {\bibfnamefont {J.~P.}\ \bibnamefont {Keener}},\ }\bibfield  {title} {\bibinfo {title} {Waves in excitable media},\ }\href {https://doi.org/10.1137/0139043} {\bibfield  {journal} {\bibinfo  {journal} {SIAM Journal on Applied Mathematics}\ }\textbf {\bibinfo {volume} {39}},\ \bibinfo {pages} {528} (\bibinfo {year} {1980})}\BibitemShut {NoStop}%
\bibitem [{\citenamefont {Tyson}\ and\ \citenamefont {Keener}(1988)}]{TYSON1988327}%
  \BibitemOpen
  \bibfield  {author} {\bibinfo {author} {\bibfnamefont {J.~J.}\ \bibnamefont {Tyson}}\ and\ \bibinfo {author} {\bibfnamefont {J.~P.}\ \bibnamefont {Keener}},\ }\bibfield  {title} {\bibinfo {title} {Singular perturbation theory of traveling waves in excitable media (a review)},\ }\href {https://doi.org/10.1016/0167-2789(88)90062-0} {\bibfield  {journal} {\bibinfo  {journal} {Physica D: Nonlinear Phenomena}\ }\textbf {\bibinfo {volume} {32}},\ \bibinfo {pages} {327} (\bibinfo {year} {1988})}\BibitemShut {NoStop}%
\bibitem [{\citenamefont {Meron}(1992)}]{MERON19921}%
  \BibitemOpen
  \bibfield  {author} {\bibinfo {author} {\bibfnamefont {E.}~\bibnamefont {Meron}},\ }\bibfield  {title} {\bibinfo {title} {Pattern formation in excitable media},\ }\href {https://doi.org/10.1016/0370-1573(92)90098-K} {\bibfield  {journal} {\bibinfo  {journal} {Physics Reports}\ }\textbf {\bibinfo {volume} {218}},\ \bibinfo {pages} {1} (\bibinfo {year} {1992})}\BibitemShut {NoStop}%
\bibitem [{\citenamefont {Rombouts}\ and\ \citenamefont {Gelens}(2020)}]{Rombouts2020}%
  \BibitemOpen
  \bibfield  {author} {\bibinfo {author} {\bibfnamefont {J.}~\bibnamefont {Rombouts}}\ and\ \bibinfo {author} {\bibfnamefont {L.}~\bibnamefont {Gelens}},\ }\bibfield  {title} {\bibinfo {title} {Synchronizing an oscillatory medium: The speed of pacemaker-generated waves},\ }\bibfield  {journal} {\bibinfo  {journal} {Physical Review Research}\ }\textbf {\bibinfo {volume} {2}},\ \href {https://doi.org/10.1103/PhysRevResearch.2.043038} {10.1103/PhysRevResearch.2.043038} (\bibinfo {year} {2020})\BibitemShut {NoStop}%
\bibitem [{\citenamefont {Keener}\ and\ \citenamefont {Tyson}(1986)}]{KEENER1986307}%
  \BibitemOpen
  \bibfield  {author} {\bibinfo {author} {\bibfnamefont {J.~P.}\ \bibnamefont {Keener}}\ and\ \bibinfo {author} {\bibfnamefont {J.~J.}\ \bibnamefont {Tyson}},\ }\bibfield  {title} {\bibinfo {title} {Spiral waves in the {B}elousov-{Z}habotinskii reaction},\ }\href {https://doi.org/10.1016/0167-2789(86)90007-2} {\bibfield  {journal} {\bibinfo  {journal} {Physica D: Nonlinear Phenomena}\ }\textbf {\bibinfo {volume} {21}},\ \bibinfo {pages} {307} (\bibinfo {year} {1986})}\BibitemShut {NoStop}%
\bibitem [{\citenamefont {Cebrián-Lacasa}\ \emph {et~al.}(2024)\citenamefont {Cebrián-Lacasa}, \citenamefont {Parra-Rivas}, \citenamefont {Ruiz-Reynés},\ and\ \citenamefont {Gelens}}]{cebrian2024six}%
  \BibitemOpen
  \bibfield  {author} {\bibinfo {author} {\bibfnamefont {D.}~\bibnamefont {Cebrián-Lacasa}}, \bibinfo {author} {\bibfnamefont {P.}~\bibnamefont {Parra-Rivas}}, \bibinfo {author} {\bibfnamefont {D.}~\bibnamefont {Ruiz-Reynés}},\ and\ \bibinfo {author} {\bibfnamefont {L.}~\bibnamefont {Gelens}},\ }\bibfield  {title} {\bibinfo {title} {Six decades of the {F}itz{H}ugh–{N}agumo model: A guide through its spatio-temporal dynamics and influence across disciplines},\ }\href {https://doi.org/https://doi.org/10.1016/j.physrep.2024.09.014} {\bibfield  {journal} {\bibinfo  {journal} {Physics Reports}\ }\textbf {\bibinfo {volume} {1096}},\ \bibinfo {pages} {1} (\bibinfo {year} {2024})}\BibitemShut {NoStop}%
\bibitem [{\citenamefont {Tyson}\ \emph {et~al.}(1989)\citenamefont {Tyson}, \citenamefont {Alexander}, \citenamefont {Manoranjan},\ and\ \citenamefont {Murray}}]{TYSON1989193}%
  \BibitemOpen
  \bibfield  {author} {\bibinfo {author} {\bibfnamefont {J.~J.}\ \bibnamefont {Tyson}}, \bibinfo {author} {\bibfnamefont {K.~A.}\ \bibnamefont {Alexander}}, \bibinfo {author} {\bibfnamefont {V.}~\bibnamefont {Manoranjan}},\ and\ \bibinfo {author} {\bibfnamefont {J.}~\bibnamefont {Murray}},\ }\bibfield  {title} {\bibinfo {title} {Spiral waves of cyclic {AMP} in a model of slime mold aggregation},\ }\href {https://doi.org/10.1016/0167-2789(89)90234-0} {\bibfield  {journal} {\bibinfo  {journal} {Physica D: Nonlinear Phenomena}\ }\textbf {\bibinfo {volume} {34}},\ \bibinfo {pages} {193} (\bibinfo {year} {1989})}\BibitemShut {NoStop}%
\bibitem [{\citenamefont {Lechleiter}\ \emph {et~al.}(1991)\citenamefont {Lechleiter}, \citenamefont {Girard}, \citenamefont {Peralta},\ and\ \citenamefont {Clapham}}]{Lechleiter1991}%
  \BibitemOpen
  \bibfield  {author} {\bibinfo {author} {\bibfnamefont {J.}~\bibnamefont {Lechleiter}}, \bibinfo {author} {\bibfnamefont {S.}~\bibnamefont {Girard}}, \bibinfo {author} {\bibfnamefont {E.}~\bibnamefont {Peralta}},\ and\ \bibinfo {author} {\bibfnamefont {D.}~\bibnamefont {Clapham}},\ }\bibfield  {title} {\bibinfo {title} {Spiral calcium wave propagation and annihilation in \textit{Xenopus laevis} oocytes},\ }\href {https://doi.org/10.1126/science.2011747} {\bibfield  {journal} {\bibinfo  {journal} {Science}\ }\textbf {\bibinfo {volume} {252}},\ \bibinfo {pages} {123} (\bibinfo {year} {1991})}\BibitemShut {NoStop}%
\bibitem [{\citenamefont {Siegert}\ and\ \citenamefont {Weijer}(1995)}]{Siegert1995}%
  \BibitemOpen
  \bibfield  {author} {\bibinfo {author} {\bibfnamefont {F.}~\bibnamefont {Siegert}}\ and\ \bibinfo {author} {\bibfnamefont {C.~J.}\ \bibnamefont {Weijer}},\ }\bibfield  {title} {\bibinfo {title} {Spiral and concentric waves organize multicellular dictyostelium mounds},\ }\href {https://doi.org/10.1016/S0960-9822(95)00184-9} {\bibfield  {journal} {\bibinfo  {journal} {Current Biology}\ }\textbf {\bibinfo {volume} {5}},\ \bibinfo {pages} {937} (\bibinfo {year} {1995})}\BibitemShut {NoStop}%
\bibitem [{\citenamefont {Ford}\ \emph {et~al.}(2023)\citenamefont {Ford}, \citenamefont {Manhart},\ and\ \citenamefont {Chubb}}]{Ford2023}%
  \BibitemOpen
  \bibfield  {author} {\bibinfo {author} {\bibfnamefont {H.~Z.}\ \bibnamefont {Ford}}, \bibinfo {author} {\bibfnamefont {A.}~\bibnamefont {Manhart}},\ and\ \bibinfo {author} {\bibfnamefont {J.~R.}\ \bibnamefont {Chubb}},\ }\bibfield  {title} {\bibinfo {title} {Controlling periodic long-range signalling to drive a morphogenetic transition},\ }\href {https://doi.org/10.7554/eLife.83796} {\bibfield  {journal} {\bibinfo  {journal} {eLife}\ ,\ \bibinfo {pages} {e83796}} (\bibinfo {year} {2023})}\BibitemShut {NoStop}%
\bibitem [{\citenamefont {Pomerening}\ \emph {et~al.}(2003)\citenamefont {Pomerening}, \citenamefont {Sontag},\ and\ \citenamefont {Ferrell~Jr}}]{Pomerening2003-ef}%
  \BibitemOpen
  \bibfield  {author} {\bibinfo {author} {\bibfnamefont {J.~R.}\ \bibnamefont {Pomerening}}, \bibinfo {author} {\bibfnamefont {E.~D.}\ \bibnamefont {Sontag}},\ and\ \bibinfo {author} {\bibfnamefont {J.~E.}\ \bibnamefont {Ferrell~Jr}},\ }\bibfield  {title} {\bibinfo {title} {Building a cell cycle oscillator: hysteresis and bistability in the activation of {C}dc2},\ }\href {https://doi.org/10.1038/ncb954} {\bibfield  {journal} {\bibinfo  {journal} {Nature cell biology}\ }\textbf {\bibinfo {volume} {5}},\ \bibinfo {pages} {346} (\bibinfo {year} {2003})}\BibitemShut {NoStop}%
\bibitem [{\citenamefont {Maryu}\ and\ \citenamefont {Yang}(2022)}]{Maryu2022-ul}%
  \BibitemOpen
  \bibfield  {author} {\bibinfo {author} {\bibfnamefont {G.}~\bibnamefont {Maryu}}\ and\ \bibinfo {author} {\bibfnamefont {Q.}~\bibnamefont {Yang}},\ }\bibfield  {title} {\bibinfo {title} {Nuclear-cytoplasmic compartmentalization of {C}yclin {B}1-{C}dk1 promotes robust timing of mitotic events},\ }\bibfield  {journal} {\bibinfo  {journal} {Cell Reports}\ }\textbf {\bibinfo {volume} {41}},\ \href {https://doi.org/10.1016/j.celrep.2022.111870} {10.1016/j.celrep.2022.111870} (\bibinfo {year} {2022})\BibitemShut {NoStop}%
\bibitem [{\citenamefont {Chang}\ and\ \citenamefont {Ferrell~Jr}(2013)}]{Chang2013}%
  \BibitemOpen
  \bibfield  {author} {\bibinfo {author} {\bibfnamefont {J.~B.}\ \bibnamefont {Chang}}\ and\ \bibinfo {author} {\bibfnamefont {J.~E.}\ \bibnamefont {Ferrell~Jr}},\ }\bibfield  {title} {\bibinfo {title} {Mitotic trigger waves and the spatial coordination of the \textit{{X}enopus} cell cycle},\ }\href {https://doi.org/10.1038/nature12321} {\bibfield  {journal} {\bibinfo  {journal} {Nature}\ }\textbf {\bibinfo {volume} {500}},\ \bibinfo {pages} {603} (\bibinfo {year} {2013})}\BibitemShut {NoStop}%
\bibitem [{\citenamefont {Puls}\ \emph {et~al.}(2024)\citenamefont {Puls}, \citenamefont {Ruiz-Reyn{\'e}s}, \citenamefont {Tavella}, \citenamefont {Jin}, \citenamefont {Kim}, \citenamefont {Gelens},\ and\ \citenamefont {Yang}}]{Puls2024}%
  \BibitemOpen
  \bibfield  {author} {\bibinfo {author} {\bibfnamefont {O.}~\bibnamefont {Puls}}, \bibinfo {author} {\bibfnamefont {D.}~\bibnamefont {Ruiz-Reyn{\'e}s}}, \bibinfo {author} {\bibfnamefont {F.}~\bibnamefont {Tavella}}, \bibinfo {author} {\bibfnamefont {M.}~\bibnamefont {Jin}}, \bibinfo {author} {\bibfnamefont {Y.}~\bibnamefont {Kim}}, \bibinfo {author} {\bibfnamefont {L.}~\bibnamefont {Gelens}},\ and\ \bibinfo {author} {\bibfnamefont {Q.}~\bibnamefont {Yang}},\ }\bibfield  {title} {\bibinfo {title} {Mitotic waves in frog egg extracts: Transition from phase waves to trigger waves},\ }\bibfield  {journal} {\bibinfo  {journal} {bioRxiv}\ }\href {https://doi.org/10.1101/2024.01.18.576267} {10.1101/2024.01.18.576267} (\bibinfo {year} {2024})\BibitemShut {NoStop}%
\bibitem [{\citenamefont {Fontanilla}\ and\ \citenamefont {Nuccitelli}(1998)}]{fontanilla1998characterization}%
  \BibitemOpen
  \bibfield  {author} {\bibinfo {author} {\bibfnamefont {R.~A.}\ \bibnamefont {Fontanilla}}\ and\ \bibinfo {author} {\bibfnamefont {R.}~\bibnamefont {Nuccitelli}},\ }\bibfield  {title} {\bibinfo {title} {Characterization of the sperm-induced calcium wave in {X}enopus eggs using confocal microscopy},\ }\href {https://doi.org/10.1016/S0006-3495(98)77650-7} {\bibfield  {journal} {\bibinfo  {journal} {Biophysical journal}\ }\textbf {\bibinfo {volume} {75}},\ \bibinfo {pages} {2079} (\bibinfo {year} {1998})}\BibitemShut {NoStop}%
\bibitem [{\citenamefont {Hara}\ \emph {et~al.}(1980)\citenamefont {Hara}, \citenamefont {Tydeman},\ and\ \citenamefont {Kirschner}}]{hara1980cytoplasmic}%
  \BibitemOpen
  \bibfield  {author} {\bibinfo {author} {\bibfnamefont {K.}~\bibnamefont {Hara}}, \bibinfo {author} {\bibfnamefont {P.}~\bibnamefont {Tydeman}},\ and\ \bibinfo {author} {\bibfnamefont {M.}~\bibnamefont {Kirschner}},\ }\bibfield  {title} {\bibinfo {title} {A cytoplasmic clock with the same period as the division cycle in {X}enopus eggs.},\ }\href {https://doi.org/10.1073/pnas.77.1.462} {\bibfield  {journal} {\bibinfo  {journal} {Proceedings of the National Academy of Sciences}\ }\textbf {\bibinfo {volume} {77}},\ \bibinfo {pages} {462} (\bibinfo {year} {1980})}\BibitemShut {NoStop}%
\bibitem [{\citenamefont {Nolet}\ \emph {et~al.}(2020)\citenamefont {Nolet}, \citenamefont {Vandervelde}, \citenamefont {Vanderbeke}, \citenamefont {Piñeros}, \citenamefont {Chang},\ and\ \citenamefont {Gelens}}]{Nolet2020}%
  \BibitemOpen
  \bibfield  {author} {\bibinfo {author} {\bibfnamefont {F.~E.}\ \bibnamefont {Nolet}}, \bibinfo {author} {\bibfnamefont {A.}~\bibnamefont {Vandervelde}}, \bibinfo {author} {\bibfnamefont {A.}~\bibnamefont {Vanderbeke}}, \bibinfo {author} {\bibfnamefont {L.}~\bibnamefont {Piñeros}}, \bibinfo {author} {\bibfnamefont {J.~B.}\ \bibnamefont {Chang}},\ and\ \bibinfo {author} {\bibfnamefont {L.}~\bibnamefont {Gelens}},\ }\bibfield  {title} {\bibinfo {title} {Nuclei determine the spatial origin of mitotic waves},\ }\bibfield  {journal} {\bibinfo  {journal} {eLife}\ }\textbf {\bibinfo {volume} {9}},\ \href {https://doi.org/10.7554/eLife.52868} {10.7554/eLife.52868} (\bibinfo {year} {2020})\BibitemShut {NoStop}%
\bibitem [{\citenamefont {Afanzar}\ \emph {et~al.}(2020)\citenamefont {Afanzar}, \citenamefont {Buss}, \citenamefont {Stearns},\ and\ \citenamefont {Ferrell}}]{Afanzar2020}%
  \BibitemOpen
  \bibfield  {author} {\bibinfo {author} {\bibfnamefont {O.}~\bibnamefont {Afanzar}}, \bibinfo {author} {\bibfnamefont {G.~K.}\ \bibnamefont {Buss}}, \bibinfo {author} {\bibfnamefont {T.}~\bibnamefont {Stearns}},\ and\ \bibinfo {author} {\bibfnamefont {J.}~\bibnamefont {Ferrell}, \bibfnamefont {James~E}},\ }\bibfield  {title} {\bibinfo {title} {The nucleus serves as the pacemaker for the cell cycle},\ }\href {https://doi.org/10.7554/eLife.59989} {\bibfield  {journal} {\bibinfo  {journal} {eLife}\ }\textbf {\bibinfo {volume} {9}},\ \bibinfo {pages} {e59989} (\bibinfo {year} {2020})}\BibitemShut {NoStop}%
\bibitem [{\citenamefont {Novak}\ and\ \citenamefont {Tyson}(1993)}]{Novak1993}%
  \BibitemOpen
  \bibfield  {author} {\bibinfo {author} {\bibfnamefont {B.}~\bibnamefont {Novak}}\ and\ \bibinfo {author} {\bibfnamefont {J.~J.}\ \bibnamefont {Tyson}},\ }\bibfield  {title} {\bibinfo {title} {{Numerical analysis of a comprehensive model of M-phase control in \textit{{X}enopus} oocyte extracts and intact embryos}},\ }\href {https://doi.org/10.1242/jcs.106.4.1153} {\bibfield  {journal} {\bibinfo  {journal} {Journal of Cell Science}\ }\textbf {\bibinfo {volume} {106}},\ \bibinfo {pages} {1153} (\bibinfo {year} {1993})}\BibitemShut {NoStop}%
\bibitem [{\citenamefont {Parra-Rivas}\ \emph {et~al.}(2023)\citenamefont {Parra-Rivas}, \citenamefont {Ruiz-Reyn{\'e}s},\ and\ \citenamefont {Gelens}}]{parra2023cell}%
  \BibitemOpen
  \bibfield  {author} {\bibinfo {author} {\bibfnamefont {P.}~\bibnamefont {Parra-Rivas}}, \bibinfo {author} {\bibfnamefont {D.}~\bibnamefont {Ruiz-Reyn{\'e}s}},\ and\ \bibinfo {author} {\bibfnamefont {L.}~\bibnamefont {Gelens}},\ }\bibfield  {title} {\bibinfo {title} {Cell cycle oscillations driven by two interlinked bistable switches},\ }\href {https://doi.org/10.1091/mbc.E22-11-0527} {\bibfield  {journal} {\bibinfo  {journal} {Molecular Biology of the Cell}\ }\textbf {\bibinfo {volume} {34}},\ \bibinfo {pages} {ar56} (\bibinfo {year} {2023})}\BibitemShut {NoStop}%
\bibitem [{\citenamefont {FitzHugh}(1961)}]{FitzHugh1961}%
  \BibitemOpen
  \bibfield  {author} {\bibinfo {author} {\bibfnamefont {R.}~\bibnamefont {FitzHugh}},\ }\bibfield  {title} {\bibinfo {title} {Impulses and physiological states in theoretical models of nerve membrane},\ }\href {https://doi.org/10.1016/S0006-3495(61)86902-6} {\bibfield  {journal} {\bibinfo  {journal} {Biophysical Journal}\ }\textbf {\bibinfo {volume} {1}},\ \bibinfo {pages} {445} (\bibinfo {year} {1961})}\BibitemShut {NoStop}%
\bibitem [{\citenamefont {Nagumo}\ \emph {et~al.}(1962)\citenamefont {Nagumo}, \citenamefont {Arimoto},\ and\ \citenamefont {Yoshizawa}}]{Nagumo1962}%
  \BibitemOpen
  \bibfield  {author} {\bibinfo {author} {\bibfnamefont {J.}~\bibnamefont {Nagumo}}, \bibinfo {author} {\bibfnamefont {S.}~\bibnamefont {Arimoto}},\ and\ \bibinfo {author} {\bibfnamefont {S.}~\bibnamefont {Yoshizawa}},\ }\bibfield  {title} {\bibinfo {title} {An active pulse transmission line simulating nerve axon},\ }\href {https://doi.org/10.1109/JRPROC.1962.288235} {\bibfield  {journal} {\bibinfo  {journal} {Proceedings of the IRE}\ }\textbf {\bibinfo {volume} {50}},\ \bibinfo {pages} {2061} (\bibinfo {year} {1962})}\BibitemShut {NoStop}%
\bibitem [{\citenamefont {Pi{\~n}eros}\ \emph {et~al.}(2024)\citenamefont {Pi{\~n}eros}, \citenamefont {Frolov}, \citenamefont {Ruiz-Reyn{\'e}s}, \citenamefont {Van~Eynde}, \citenamefont {Cavin-Meza}, \citenamefont {Heald},\ and\ \citenamefont {Gelens}}]{Pineros2024.07.28.605512}%
  \BibitemOpen
  \bibfield  {author} {\bibinfo {author} {\bibfnamefont {L.}~\bibnamefont {Pi{\~n}eros}}, \bibinfo {author} {\bibfnamefont {N.}~\bibnamefont {Frolov}}, \bibinfo {author} {\bibfnamefont {D.}~\bibnamefont {Ruiz-Reyn{\'e}s}}, \bibinfo {author} {\bibfnamefont {A.}~\bibnamefont {Van~Eynde}}, \bibinfo {author} {\bibfnamefont {G.}~\bibnamefont {Cavin-Meza}}, \bibinfo {author} {\bibfnamefont {R.}~\bibnamefont {Heald}},\ and\ \bibinfo {author} {\bibfnamefont {L.}~\bibnamefont {Gelens}},\ }\bibfield  {title} {\bibinfo {title} {The nuclear-cytoplasmic ratio controls the cell cycle period in compartmentalized frog egg extract},\ }\bibfield  {journal} {\bibinfo  {journal} {bioRxiv}\ }\href {https://doi.org/10.1101/2024.07.28.605512} {10.1101/2024.07.28.605512} (\bibinfo {year} {2024})\BibitemShut {NoStop}%
\bibitem [{\citenamefont {Anderson}\ \emph {et~al.}(2017)\citenamefont {Anderson}, \citenamefont {Gelens}, \citenamefont {Baker},\ and\ \citenamefont {Ferrell}}]{anderson2017desynchronizing}%
  \BibitemOpen
  \bibfield  {author} {\bibinfo {author} {\bibfnamefont {G.~A.}\ \bibnamefont {Anderson}}, \bibinfo {author} {\bibfnamefont {L.}~\bibnamefont {Gelens}}, \bibinfo {author} {\bibfnamefont {J.~C.}\ \bibnamefont {Baker}},\ and\ \bibinfo {author} {\bibfnamefont {J.~E.}\ \bibnamefont {Ferrell}},\ }\bibfield  {title} {\bibinfo {title} {Desynchronizing embryonic cell division waves reveals the robustness of \textit{{X}enopus laevis} development},\ }\href {https://doi.org/10.1016/j.celrep.2017.09.017} {\bibfield  {journal} {\bibinfo  {journal} {Cell reports}\ }\textbf {\bibinfo {volume} {21}},\ \bibinfo {pages} {37} (\bibinfo {year} {2017})}\BibitemShut {NoStop}%
\bibitem [{\citenamefont {Johmura}\ \emph {et~al.}(2011)\citenamefont {Johmura}, \citenamefont {Soung}, \citenamefont {Park}, \citenamefont {Yu}, \citenamefont {Zhou}, \citenamefont {Bang}, \citenamefont {Kim}, \citenamefont {Veenstra}, \citenamefont {Erikson},\ and\ \citenamefont {Lee}}]{johmura2011regulation}%
  \BibitemOpen
  \bibfield  {author} {\bibinfo {author} {\bibfnamefont {Y.}~\bibnamefont {Johmura}}, \bibinfo {author} {\bibfnamefont {N.-K.}\ \bibnamefont {Soung}}, \bibinfo {author} {\bibfnamefont {J.-E.}\ \bibnamefont {Park}}, \bibinfo {author} {\bibfnamefont {L.-R.}\ \bibnamefont {Yu}}, \bibinfo {author} {\bibfnamefont {M.}~\bibnamefont {Zhou}}, \bibinfo {author} {\bibfnamefont {J.~K.}\ \bibnamefont {Bang}}, \bibinfo {author} {\bibfnamefont {B.-Y.}\ \bibnamefont {Kim}}, \bibinfo {author} {\bibfnamefont {T.~D.}\ \bibnamefont {Veenstra}}, \bibinfo {author} {\bibfnamefont {R.~L.}\ \bibnamefont {Erikson}},\ and\ \bibinfo {author} {\bibfnamefont {K.~S.}\ \bibnamefont {Lee}},\ }\bibfield  {title} {\bibinfo {title} {Regulation of microtubule-based microtubule nucleation by mammalian polo-like kinase 1},\ }\href {https://doi.org/10.1073/pnas.1106223108} {\bibfield  {journal} {\bibinfo  {journal} {Proceedings of the National Academy of Sciences}\ }\textbf {\bibinfo {volume} {108}},\ \bibinfo {pages} {11446} (\bibinfo {year}
  {2011})}\BibitemShut {NoStop}%
\bibitem [{\citenamefont {Ishihara}\ \emph {et~al.}(2016)\citenamefont {Ishihara}, \citenamefont {Korolev},\ and\ \citenamefont {Mitchison}}]{Ishihara_eLife}%
  \BibitemOpen
  \bibfield  {author} {\bibinfo {author} {\bibfnamefont {K.}~\bibnamefont {Ishihara}}, \bibinfo {author} {\bibfnamefont {K.~S.}\ \bibnamefont {Korolev}},\ and\ \bibinfo {author} {\bibfnamefont {T.~J.}\ \bibnamefont {Mitchison}},\ }\bibfield  {title} {\bibinfo {title} {Physical basis of large microtubule aster growth},\ }\href {https://doi.org/10.7554/eLife.19145} {\bibfield  {journal} {\bibinfo  {journal} {eLife}\ }\textbf {\bibinfo {volume} {5}},\ \bibinfo {pages} {e19145} (\bibinfo {year} {2016})}\BibitemShut {NoStop}%
\bibitem [{\citenamefont {Rinaldin}\ \emph {et~al.}(2024)\citenamefont {Rinaldin}, \citenamefont {Kickuth}, \citenamefont {Dalton}, \citenamefont {Xu}, \citenamefont {Di~Talia},\ and\ \citenamefont {Brugu{\'e}s}}]{Rinaldin2024-wz}%
  \BibitemOpen
  \bibfield  {author} {\bibinfo {author} {\bibfnamefont {M.}~\bibnamefont {Rinaldin}}, \bibinfo {author} {\bibfnamefont {A.}~\bibnamefont {Kickuth}}, \bibinfo {author} {\bibfnamefont {B.}~\bibnamefont {Dalton}}, \bibinfo {author} {\bibfnamefont {Y.}~\bibnamefont {Xu}}, \bibinfo {author} {\bibfnamefont {S.}~\bibnamefont {Di~Talia}},\ and\ \bibinfo {author} {\bibfnamefont {J.}~\bibnamefont {Brugu{\'e}s}},\ }\bibfield  {title} {\bibinfo {title} {Robust cytoplasmic partitioning by solving an intrinsic cytoskeletal instability},\ }\bibfield  {journal} {\bibinfo  {journal} {bioRxiv}\ }\href {https://doi.org/10.1101/2024.03.12.584684} {10.1101/2024.03.12.584684} (\bibinfo {year} {2024})\BibitemShut {NoStop}%
\bibitem [{\citenamefont {Zykov}(2018)}]{Zykov2018}%
  \BibitemOpen
  \bibfield  {author} {\bibinfo {author} {\bibfnamefont {V.~S.}\ \bibnamefont {Zykov}},\ }\bibfield  {title} {\bibinfo {title} {Spiral wave initiation in excitable media},\ }\href {https://doi.org/10.1098/rsta.2017.0379} {\bibfield  {journal} {\bibinfo  {journal} {Philosophical Transactions of the Royal Society A: Mathematical, Physical and Engineering Sciences}\ }\textbf {\bibinfo {volume} {376}},\ \bibinfo {pages} {20170379} (\bibinfo {year} {2018})}\BibitemShut {NoStop}%
\bibitem [{\citenamefont {Bub}\ \emph {et~al.}(2002)\citenamefont {Bub}, \citenamefont {Shrier},\ and\ \citenamefont {Glass}}]{Glass_2022}%
  \BibitemOpen
  \bibfield  {author} {\bibinfo {author} {\bibfnamefont {G.}~\bibnamefont {Bub}}, \bibinfo {author} {\bibfnamefont {A.}~\bibnamefont {Shrier}},\ and\ \bibinfo {author} {\bibfnamefont {L.}~\bibnamefont {Glass}},\ }\bibfield  {title} {\bibinfo {title} {Spiral wave generation in heterogeneous excitable media},\ }\href {https://doi.org/10.1103/PhysRevLett.88.058101} {\bibfield  {journal} {\bibinfo  {journal} {Phys. Rev. Lett.}\ }\textbf {\bibinfo {volume} {88}},\ \bibinfo {pages} {058101} (\bibinfo {year} {2002})}\BibitemShut {NoStop}%
\bibitem [{\citenamefont {Sepulchre}\ and\ \citenamefont {Babloyantz}(1993)}]{Babloyantz_spirals}%
  \BibitemOpen
  \bibfield  {author} {\bibinfo {author} {\bibfnamefont {J.~A.}\ \bibnamefont {Sepulchre}}\ and\ \bibinfo {author} {\bibfnamefont {A.}~\bibnamefont {Babloyantz}},\ }\bibfield  {title} {\bibinfo {title} {Motions of spiral waves in oscillatory media and in the presence of obstacles},\ }\href {https://doi.org/10.1103/PhysRevE.48.187} {\bibfield  {journal} {\bibinfo  {journal} {Phys. Rev. E}\ }\textbf {\bibinfo {volume} {48}},\ \bibinfo {pages} {187} (\bibinfo {year} {1993})}\BibitemShut {NoStop}%
\bibitem [{\citenamefont {Agladze}\ \emph {et~al.}(1994)\citenamefont {Agladze}, \citenamefont {Keener}, \citenamefont {Müller},\ and\ \citenamefont {Panfilov}}]{Panfilov_spirals}%
  \BibitemOpen
  \bibfield  {author} {\bibinfo {author} {\bibfnamefont {K.}~\bibnamefont {Agladze}}, \bibinfo {author} {\bibfnamefont {J.~P.}\ \bibnamefont {Keener}}, \bibinfo {author} {\bibfnamefont {S.~C.}\ \bibnamefont {Müller}},\ and\ \bibinfo {author} {\bibfnamefont {A.}~\bibnamefont {Panfilov}},\ }\bibfield  {title} {\bibinfo {title} {Rotating spiral waves created by geometry},\ }\href {http://www.jstor.org/stable/2883923} {\bibfield  {journal} {\bibinfo  {journal} {Science}\ }\textbf {\bibinfo {volume} {264}},\ \bibinfo {pages} {1746} (\bibinfo {year} {1994})}\BibitemShut {NoStop}%
\bibitem [{\citenamefont {Rombouts}\ and\ \citenamefont {Gelens}(2021)}]{Rombouts_PhysRevE.104.014220}%
  \BibitemOpen
  \bibfield  {author} {\bibinfo {author} {\bibfnamefont {J.}~\bibnamefont {Rombouts}}\ and\ \bibinfo {author} {\bibfnamefont {L.}~\bibnamefont {Gelens}},\ }\bibfield  {title} {\bibinfo {title} {Analytical approximations for the speed of pacemaker-generated waves},\ }\href {https://doi.org/10.1103/PhysRevE.104.014220} {\bibfield  {journal} {\bibinfo  {journal} {Phys. Rev. E}\ }\textbf {\bibinfo {volume} {104}},\ \bibinfo {pages} {014220} (\bibinfo {year} {2021})}\BibitemShut {NoStop}%
\bibitem [{\citenamefont {Lee}(1997)}]{PhysRevLett.79.2907}%
  \BibitemOpen
  \bibfield  {author} {\bibinfo {author} {\bibfnamefont {K.~J.}\ \bibnamefont {Lee}},\ }\bibfield  {title} {\bibinfo {title} {Wave pattern selection in an excitable system},\ }\href {https://doi.org/10.1103/PhysRevLett.79.2907} {\bibfield  {journal} {\bibinfo  {journal} {Phys. Rev. Lett.}\ }\textbf {\bibinfo {volume} {79}},\ \bibinfo {pages} {2907} (\bibinfo {year} {1997})}\BibitemShut {NoStop}%
\bibitem [{\citenamefont {Bement}\ \emph {et~al.}(2015)\citenamefont {Bement}, \citenamefont {Leda}, \citenamefont {Moe}, \citenamefont {Kita}, \citenamefont {Larson}, \citenamefont {Golding}, \citenamefont {Pfeuti}, \citenamefont {Su}, \citenamefont {Miller}, \citenamefont {Goryachev} \emph {et~al.}}]{bement2015activator}%
  \BibitemOpen
  \bibfield  {author} {\bibinfo {author} {\bibfnamefont {W.~M.}\ \bibnamefont {Bement}}, \bibinfo {author} {\bibfnamefont {M.}~\bibnamefont {Leda}}, \bibinfo {author} {\bibfnamefont {A.~M.}\ \bibnamefont {Moe}}, \bibinfo {author} {\bibfnamefont {A.~M.}\ \bibnamefont {Kita}}, \bibinfo {author} {\bibfnamefont {M.~E.}\ \bibnamefont {Larson}}, \bibinfo {author} {\bibfnamefont {A.~E.}\ \bibnamefont {Golding}}, \bibinfo {author} {\bibfnamefont {C.}~\bibnamefont {Pfeuti}}, \bibinfo {author} {\bibfnamefont {K.-C.}\ \bibnamefont {Su}}, \bibinfo {author} {\bibfnamefont {A.~L.}\ \bibnamefont {Miller}}, \bibinfo {author} {\bibfnamefont {A.~B.}\ \bibnamefont {Goryachev}}, \emph {et~al.},\ }\bibfield  {title} {\bibinfo {title} {Activator--inhibitor coupling between {R}ho signalling and actin assembly makes the cell cortex an excitable medium},\ }\href {https://doi.org/10.1038/ncb3251} {\bibfield  {journal} {\bibinfo  {journal} {Nature cell biology}\ }\textbf {\bibinfo {volume} {17}},\ \bibinfo {pages} {1471} (\bibinfo
  {year} {2015})}\BibitemShut {NoStop}%
\bibitem [{\citenamefont {Michaud}\ \emph {et~al.}(2022)\citenamefont {Michaud}, \citenamefont {Leda}, \citenamefont {Swider}, \citenamefont {Kim}, \citenamefont {He}, \citenamefont {Landino}, \citenamefont {Valley}, \citenamefont {Huisken}, \citenamefont {Goryachev}, \citenamefont {von Dassow} \emph {et~al.}}]{bement2024patterning}%
  \BibitemOpen
  \bibfield  {author} {\bibinfo {author} {\bibfnamefont {A.}~\bibnamefont {Michaud}}, \bibinfo {author} {\bibfnamefont {M.}~\bibnamefont {Leda}}, \bibinfo {author} {\bibfnamefont {Z.~T.}\ \bibnamefont {Swider}}, \bibinfo {author} {\bibfnamefont {S.}~\bibnamefont {Kim}}, \bibinfo {author} {\bibfnamefont {J.}~\bibnamefont {He}}, \bibinfo {author} {\bibfnamefont {J.}~\bibnamefont {Landino}}, \bibinfo {author} {\bibfnamefont {J.~R.}\ \bibnamefont {Valley}}, \bibinfo {author} {\bibfnamefont {J.}~\bibnamefont {Huisken}}, \bibinfo {author} {\bibfnamefont {A.~B.}\ \bibnamefont {Goryachev}}, \bibinfo {author} {\bibfnamefont {G.}~\bibnamefont {von Dassow}}, \emph {et~al.},\ }\bibfield  {title} {\bibinfo {title} {A versatile cortical pattern-forming circuit based on {R}ho, {F}-actin, {E}ct2, and {RGA}-3/4},\ }\href {https://doi.org/10.1083/jcb.202203017} {\bibfield  {journal} {\bibinfo  {journal} {Journal of Cell Biology}\ }\textbf {\bibinfo {volume} {221}},\ \bibinfo {pages} {e202203017} (\bibinfo {year}
  {2022})}\BibitemShut {NoStop}%
\bibitem [{\citenamefont {Gelens}\ \emph {et~al.}(2015)\citenamefont {Gelens}, \citenamefont {Huang},\ and\ \citenamefont {Ferrell}}]{GELENS2015892}%
  \BibitemOpen
  \bibfield  {author} {\bibinfo {author} {\bibfnamefont {L.}~\bibnamefont {Gelens}}, \bibinfo {author} {\bibfnamefont {K.}~\bibnamefont {Huang}},\ and\ \bibinfo {author} {\bibfnamefont {J.}~\bibnamefont {Ferrell}},\ }\bibfield  {title} {\bibinfo {title} {How does the \textit{{X}enopus laevis} embryonic cell cycle avoid spatial chaos?},\ }\href {https://doi.org/10.1016/j.celrep.2015.06.070} {\bibfield  {journal} {\bibinfo  {journal} {Cell Reports}\ }\textbf {\bibinfo {volume} {12}},\ \bibinfo {pages} {892} (\bibinfo {year} {2015})}\BibitemShut {NoStop}%
\end{thebibliography}%

\end{document}